\documentstyle[preprint,aps,epsfig]{revtex}
\input epsf

\begin{document}
\draft
\tighten

\oddsidemargin -0.5cm \evensidemargin -0.5cm
\newcommand{\be}{\begin{eqnarray}}
\newcommand{\ee}{\end{eqnarray}}

\title{Probing halo nucleus structure through intermediate energy elastic scattering.}

\author{R. Crespo  \footnote{E-mail:raquel@wotan.ist.utl.pt}     }
\address{ Departamento de
F\'{\i}sica, Instituto Superior T\'ecnico, Lisboa, Portugal}

\author{R.C. Johnson  \footnote{E-mail:R.Johnson@surrey.ac.uk }      }
\address{ Department of Physics, University of Surrey,
Guildford,\\ Surrey, GU2 5XH, United Kingdom }

\date{\today}

\maketitle

\begin{abstract}
This work addresses the question of precisely what features of few body 
models of halo nuclei are probed by elastic scattering on protons at high  centre-of-mass energies.
Our treatment is based on a multiple scattering expansion of the proton-projectile transition amplitude 
in a form which is well adapted to the weakly bound cluster picture of halo nuclei.
In the specific case of $^{11}$Li scattering from protons at 800 MeV/u we show that  because core recoil effects are significant, scattering crosssections can not, in general,  be deduced from knowledge of the total matter density alone. 

We advocate that the optical potential concept for the scattering of
halo nuclei on protons should be avoided and that the multiple scattering
series for the full transition amplitude should be used instead.

\end{abstract}

\medskip
\centerline{[PACS catagories: 24.10.--i, 24.10.Ht, 25.40.Cm]}

\vfill
\eject

\section{Introduction}

Models of light halo nuclei have been developed \cite{Zhukov,mafia2,ianwf} where the few
body degrees of freedom of a system of loosely bound valence nucleons orbiting
around a relatively tightly bound core are properly taken into account.
Short range, center of mass and some Pauli principle
effects are often included in these models.

In this work we develop a multiple scattering expansion of the nucleon-projectile transition amplitude for proton scattering from a few
body system. When the projectile is
composed of weakly bound sub-systems a multiple scattering expansion of the 
transition amplitude in terms of 2-body t-matrices describing proton scattering from the projectile sub-systems is expected to converge rapidly \cite{watson}. 
The elastic scattering observables may then be derived directly from this    expansion. We contrast this with our
earlier work \cite{CJT1,CrespoLi9} which is based on a multiple scattering expansion of the 
optical model operator and therefore treats the ground and excited states of the projectile
on a different footing. The present approach is more appropriate for few-body 
projectiles at high projectile energy.

Our aim in this work is to understand the nuclear structure features that
should be incorporated into the reaction mechanism in order to describe elastic
scattering of halo nuclei from stable nuclei. In particular it is of
considerable interest to examine how far elastic scattering observables
probe correlation effects among projectile nucleons \cite{jim,alkhazov}.

\section{Multiple scattering expansion}

We consider the transition amplitude, $T$, for scattering of a proton from a
many body-system composed of a small number of sub-systems. We have in mind, for example, $^{11}$Li
 assumed to be well described by two valence loosely bound nucleons
orbiting around a $^{9}$Li core. $T$ can be written as a multiple
scattering expansion in the transition amplitudes $\hat{t}_{\cal I}$ for proton
scattering from each projectile sub-system ${\cal I}$ \cite{watson}. We ignore explicit reference to excitations of the sub-systems, although each $\hat{t}_{\cal I}$ may implicitly contain  effects due to such excitations and will certainly do so if, as we shall assume, they describe elastic proton-sub-system scattering. In other words our model assumes that we only need refer explicitly to excitations of the projectile which involve changes in the relative motion of the sub-systems in the projectile. This is consistent with standard few-body treatments of reactions involving halo nuclei \cite{review,johnson1}. 

The multiple
scattering expansion can be written 
 \be
T = \sum_{\cal I}\hat{t}_{\cal I}  +
 \sum_{\cal I} \hat{t}_{\cal I} G_0 \sum_{{\cal J} \neq {\cal I}}
  \hat{t}_{\cal J} + \cdots
\label{TMSexpa}
\ee
where the proton - ${\cal I}$ subsystem transition amplitude satisfies
\be
\hat{t}_{\cal I} = v_{\cal I} + v_{\cal I} G_0 \hat{t}_{\cal I} ~~.
\ee
The propagator $G_0$ contains the kinetic energy operators of the
projectile
and all the target subsystems,
$G_0 = \left( E^+ - K \right)^{-1}$.
Here E is the kinetic energy, $E = \frac{\hbar^2 k_i^2}{2 \mu_{NA}}$ in the
overall center of mass
frame, and  $\mu_{NA}$ is the proton-projectile reduced mass. We ignore the interaction between projectile sub-systems in $G_0$ (impulse approximation).
We note that in the multiple scattering expansion eq.(\ref{TMSexpa})
both elastic and inelastic excitations of the relative motion of the subsystems in intermediate states are taken into account. For proton scattering from halo nuclei
the inelastic channels associated with breakup of the halo nucleus into its sub-systems are expected to contribute
significantly to the transition amplitude.

   In this paper we truncate the series in eq.(\ref{TMSexpa}) at the 
double scattering terms. We have not evaluated third order terms and we 
do not claim that they are negligible. They could be handled using the
 techniques of, for example, reference \cite{Garcilazo}. Our purpose
 here is to assess the applicability of the standard approach for proton 
scattering on light nuclei. We will show that inadequacies show up
 even at the second order level.

A second aim of our work is to understand the role of various types of 
correlations in elastic scattering from halo systems. In this paper we 
make a numerical study of the case of  proton scattering from a $^{11}$Li
 projectile at intermediate energies. Our formalism could also be applied 
to p-$^{6}$He scattering which has been studied extensively elsewhere 
using methods which do not use a truncated multiple scattering 
expansion \cite{jim2} but do not lend themselves well to delineating 
the role of correlations in an explicit way.    

We assume that the projectile wave function  can   be
written as the product of the core internal wave function
$\varphi_{_{ C} }$ and the wave function of the two body valence system
 relative to the core $\varphi_{_{nn}}(\vec{r},\vec{R})$,
where $\vec{r}=\vec{r}_2-\vec{r}_3$ is the relative position of the two
 valence bodies 2 and 3, and 
$\vec{R}$ is the vector from the core centre of mass (particle 4) to the centre of mass of the valence pair.

For projectile energies in the intermediate energy region the
relative momentum between each subsystem pair is small in comparison
with the projectile momentum and will be neglected  
wherever it appears. The elastic transition amplitude to second
order in the proton-subsystem transition amplitudes, involves 
single scattering terms where the projectile scatters from each
target subsystem and double scattering terms where the proton
scatters from one subsystem and rescatters from another.

\subsection{Single scattering}

The contribution to the single scattering term from proton  scattering  from one of the valence particles, for
example particle 2,  is given by
\be
\langle  \vec{k}_f  \Phi |  \hat{t}_{12} | \vec{k}_i  \Phi \rangle
= \langle  \vec{k}_f  \varphi_{nn} |  \hat{t}_{12}
 | \vec{k}_i  \varphi_{nn} \rangle  =
\hat{t}_{12}(\omega_{12}, \vec{\Delta}) \rho_{v}(\vec{\Delta}) \label{t12}
\ee
where $\rho_{v}(\vec{\Delta})$ is defined in terms of the 2-body halo density
\be
\rho_{2}(\vec{\Delta}_{1},\vec{\Delta}_{2}) =
\int d\vec{Q}_1 d\vec{Q}_2 \,
\varphi^*_{nn}(\vec{Q}_1,\vec{Q}_2)
 \varphi_{nn}(\vec{Q}_1 +  \vec{\Delta}_1,
\vec{Q}_2 +  \vec{\Delta}_2)~~ \label{rho2},
\ee
by
\be
\rho_{v}(\vec{\Delta})=\rho_{2}({\scriptstyle \frac{m_3}{M_{23}}} \vec{\Delta}, {\scriptstyle  \frac{m_4}{M_{234}}} \vec{\Delta})~~\label{rhov},
\ee
where $M_{23}=m_{2}+m_{3}, M_{234}=m_{2}+m_{3}+m_{4}$, etc.
In eq.(\ref{rho2}) $\varphi_{nn}(\vec{Q}_1,\vec{Q}_2)$ is the 
Fourier transform of wave function of the two body valence system relative
to the core $\varphi_{_{nn}}(\vec{r},\vec{R})$.
In the case $m_2=m_3=m_{n}$ the quantity $\rho_{v}(\vec{\Delta})$ is just the Fourier transform of the probability density
$\rho(\vec{x})$ of finding a valence neutron at a distance 
${\vec x}$ from the center of mass of the  projectile as defined by Zhukov {\em et al} \cite{Zhukov}.

The energy parameter $\omega_{12}$ in eq.(\ref{t12}) is given by
\be
\omega_{12}= E \left[1-\frac{m_{1}M_{34}}{M_{12}M_{234}}\right]
\label{w12}
\ee
and reduces to $\omega_{12} = E/2$ in the limit of
 $m_4  \gg m_3, m_2 $.

The contribution to the single scattering term from proton scattering from the core is
\be
\langle \vec{k}_f \Phi | \hat{t}_{14} |\vec{k}_i \, \Phi \rangle =
\langle \varphi_{\rm core} | \hat{t}_{14}(\omega_{14}, \vec{\Delta})
| \varphi_{\rm core} \rangle  \rho_2(0, {\scriptstyle  \frac{M_{23}}{M_{234}}} \vec{\Delta})
\label{t14li11}
\ee
where $\rho_2$ is defined in eq.(\ref{rho2}) and the arguments in eq.(\ref{t14li11}) mean that what is involved is the density distribution for the motion of the core
center of mass, as defined by Zhukov {\em et al}\cite{Zhukov},
\be
\rho_2(0, {\scriptstyle  \frac{M_{23}}{M_{234}}} \vec{\Delta})
 =
\int d\vec{Q}_1 d\vec{Q}_2 \,
\varphi^*_{nn}(\vec{Q}_1,\vec{Q}_2)
 \varphi_{nn}(\vec{Q}_1 ,
\vec{Q}_2 + {\scriptstyle  \frac{M_{23}}{M_{234}}} \vec{\Delta})~~ \label{rho2rc}.
\ee
and the energy parameter $\omega_{14}$ is given by
\be
\omega_{14}= E \left[1-\frac{m_{1}M_{23}}{M_{14}M_{234}}\right]
\label{w14}
\ee
In the limit of  $m_4  \gg 1$, $\omega_{14}= E$ and 
$\rho_2(0, {\scriptstyle  \frac{M_{23}}{M_{234}}} \vec{\Delta})\rightarrow \rho_2(0, 0)
 = 1$ so that eq.(\ref{t14li11}) reduces to the
expected expression for the proton scattering
from subsystem 4.
 
Within our model, there are two contributions to  the single scattering term.
Firstly a valence contribution given by the product of the projectile 
valence system transition amplitude and $\rho_v(\vec{\Delta})$.
Secondly a  core contribution in which the nucleon-core transition amplitude   is  modulated by the form factor 
$\rho_2(0, {\scriptstyle  \frac{m_{23}}{M_{234}}} \vec{\Delta})$
whose departure from unity arises from the motion of the core centre of mass
about the projectile  center of mass.
This modulation differs from standard applications of
the  multiple scattering expansion of the optical potential operator 
\cite{CrespoLi9,alkhazov} that modulate the core matter density distribution
 $\rho_{\rm C}$ by that form factor. 

The relevant halo structure information for the single scattering term
is thus  contained in the matter density form factors $\rho_{v}(\vec{\Delta})$ and
 $\rho_2(0,\, {\scriptstyle  \frac{M_{23}}{M_{234}}} \vec{\Delta})$.

\subsection{Double scattering}

We next evaluate the double scattering term in the $^{11}$Li case.
We distinguish the terms where the proton scatters from 
the valence neutrons 2 and 3 and the term where the proton scatters 
once from the core and once from a valence particle.
In the former case we find that
 \be
\langle \vec{k}_f \, \Phi  | \hat{t}_{12} G_0 \hat{t}_{13} |
\vec{k}_i\, \Phi \rangle  &=&
\langle \vec{k}_f \, \varphi_{nn}  | \hat{t}_{12} G_0 \hat{t}_{13} |
\vec{k}_i\, \varphi_{nn} \rangle   ~~~~~~~~~~~~~~~~~~~~~~~~~~ \nonumber \\
&=& \int d\vec{q} \,\,\hat{t}_{12}
\left(\omega_{12}, {\scriptstyle \frac{m_2}{M_{23}} }
\vec{\Delta} + \vec{q} \right)
\hat{t}_{13} \left(\omega_{13},
{\scriptstyle  \frac{m_3}{M_{23}}} \vec{\Delta} - \vec{q} \right)\,
G_0(\vec{q}) \rho_2(\vec{q},\,\frac{m_4}{M_{234}}\vec{\Delta}), \label{p23}
\ee
where $\rho_{2}(\vec{\Delta}_{1},\vec{\Delta}_{2})$ is defined in eq.(\ref{rho2}) and
\be
G_{0}(\vec{q})=2\frac{\mu_{1(23)}}{\hbar^{2}}\left[k_{i}^{2}-\left(\frac{(m_{3}\vec{k}_{f}+m_{2}\vec{k}_{i})}{M_{23}}+\vec{q}\right)^{2}+i\epsilon\right]^{-1} \label{g0}.
\ee

 In the case of a heavy core, 
\be
\lim_{m_4 \rightarrow \infty}
\rho_{2}(\vec{q},{\scriptstyle \frac{m_4}{M_{234}}}\vec{\Delta})& =&
{ \int d\vec{Q}_1  d\vec{Q}_2
\,  \varphi^*_{nn}} \left(\vec{Q}_1 ,\vec{Q}_2
\right)
{\varphi_{nn}} \left(\vec{Q}_1+\vec{q} ,\vec{Q}_2+\vec{\Delta} \right)
\nonumber \\
&=&  {    \int d\vec{r}_2  d\vec{r}_3
 \,\,e^{i(\frac{m_2}{M_{23}}\vec{\Delta} + \vec{q}).\vec{r}_2}
 e^{i(\frac{m_3}{M_{23}}\vec{\Delta} - \vec{q}).\vec{r}_3}  
 |{ \overline{\varphi}_{nn}}(\vec{r}_2, \vec{r}_3)|^2 } ~~,
\ee
where $ \overline{\varphi}_{nn}(\vec{r}_2,\vec{r}_3)=\varphi_{_{nn}}(\vec{r},\vec{R})$ is the valence wavefunction expressed in terms of $\vec{r}_2$ and $\vec{r}_3$, the position vectors of the 2 valence particles relative to the core. 
Therefore, this density function involves
two-body correlations among the valence particles even in the heavy core limit.

The valence system - core double scattering term is given by
\be
\langle \Phi | \hat{t}_{12} G_1 \hat{t}_{14} |
 \Phi \rangle =
 \int d\vec{q} \, &\hat{t}&_{12}
 \left(\omega_{12},
{\scriptstyle  \frac{M_{23}}{M_{234}}} \vec{\Delta} + \vec{q} \right)
\langle \varphi_{\rm core} | \hat{t}_{41} \left(\omega_{14}, 
{\scriptstyle \frac{m_4}{M_{234}} }  \vec{\Delta} - \vec{q} \right)
| \varphi_{\rm core} \rangle \nonumber \\ &G&_1(\vec{q}) 
\rho_{2}(\frac{m_3}{M_{23}}\vec{q}+\frac{m_3}{M_{234}}\vec{\Delta},\, 
\vec{q})~~. \label{t12t14li11}
\ee
where
\be
G_{1}(\vec{q})=2\frac{\mu_{1(234)}}{\hbar^{2}}\left[k_{i}^{2}-\left(\frac{(m_{4}\vec{k}_{f}+m_{23}\vec{k}_{i})}{M_{234}}+\vec{q}\right)^{2}+i\epsilon\right]^{-1} \label{g1}.
\ee
For $m_4 \gg m_2, \, m_3$ 
\be
\lim_{m_4 \rightarrow \infty}\rho_{2}(\frac{m_3}{M_{23}}\vec{q}+\frac{m_3}{M_{234}}\vec{\Delta},\, \vec{q}) = { \int d\vec{r}_2 
 e^{i( \vec{q}\cdot \vec{r}_2) } \int \, d\vec{r}_3   |{ \overline{\varphi}_{nn}}(\vec{r}_2, \vec{r}_3)|^2}\label{limDScC} \,\,\,~~.
\ee
In contrast to the double scattering term arising from the two valence particles, the particular elements of $\rho_{2}$ which enters 
in the heavy core limit is just the one body density of subsystem 2 in
 the halo.

\subsection{Numerical results for $^{11}$Li scattering   at 800 MeV/u}

In order to obtain some quantitative idea of the various terms we have
 identified, we have 
evaluated the multiple
scattering expansion for the specific case of proton scattering at 800Mev/u.
 from a 3-body model of $^{11}$Li. For the purposes of the estimate, only the central components of the transition
amplitudes were taken into account. Coulomb interaction effects were
not included. 

For the description of $^{11}$Li
we take the Fadeev wave functions of Thompson and Zhukov \cite{ianwf}
referred in that work as the P3 model. In
describing the $^{9}$Li ground state matter density distribution 
we consider a simplified structure model of a Gaussian distribution with a 
range chosen to reproduce the rms radius \cite{CrespoLi9}.
The first and second order terms were evaluated using a NN transition
amplitude derived from the Paris potential \cite{redish,paris} evaluated at the
appropriate fixed energy parameter with finite mass effects properly
taken into account. 
The transition amplitude for proton scattering from $^{9}$Li was 
generated by an optical potential calculated in the
single scattering approximation appropriate for intermediate energy elastic
scattering \cite{CJT1}.

In the evaluation of the second order terms, the  propagators
were evaluated using the eikonal approximation
and the principal value term was neglected. For example we use
\be
G_{1}(\vec{q})=-\frac{\mu_{1(234)}}{\hbar^{2}}\left(\frac{1}{\vec{k}_{i}.\vec{q}+i\epsilon}\right)~~. \label{gapprox}
\ee
 An explicit evaluation using
gaussian functional forms for the transition matrices and densities involved shows that for small scattering angles   the ratio of the principal value and delta function terms in eq.(\ref{gapprox}) is less than $1/k_{i}R$, where R is a measure of the halo size. This ratio is very small in the cases we consider.  
 
In Fig.1  we show the differential cross section for $^{11}$Li scattering 
from a proton target at  800MeV/nucleon for centre-of-mass  in the range
we expect to be covered by  experiments (e.g.  \cite{alkhazov}).
The dashed curve was evaluated from the single scattering contributions
 eqs.(\ref{t12}),(\ref{t14li11}).
The solid curve includes in addition the double scattering contributions 
valence-valence eq.(\ref{p23}) and valence-core eq.(\ref{t12t14li11}).
We emphasise that in the present context "double scattering" means 2nd order 
in the proton-subsystem $t$ matrix.
Terms of all orders in the p-subsystem potentials are included in our 1st
order terms.
The other curves in the figure are obtained by taking 
$\rho_2(0, {\scriptstyle  \frac{M_{23}}{M_{234}}} \vec{\Delta}) = 1$ 
for all $\vec{\Delta}$ in eq.(\ref{t14li11}).
This limit corresponds to ignoring  the relative motion of the core and
projectile centres of mass.
The dotted-dashed and dotted curves correspond to single and double scattering
calculated cross section respectively, and  clearly 
shows that the inclusion of the relative motion of the core and
projectile centres of mass has a significant effect in the 
calculated differential cross section.

\section{Discussion}

In the context of nucleon scattering from conventional stable heavy nuclei 
one usually associates 2-body correlation effects 
with the double scattering terms which in the present case would mean through
 the 2-body density   
$\rho_2(\vec{q},\,\frac{m_4}{M_{234}}\vec{\Delta})$ in  eq.(\ref{p23}). 
The contribution from this to the second order
term are very small here, and only valence-core double scattering
contributions remain relevant.
However, that does not mean that the scattering is sensitive only to the
projectile density $\rho_{v}(\vec{\Delta})$ of eq.(\ref{rhov}),
which we might reasonably call "the halo density". 
The scattering involves the halo wavefunction in several other distinct ways:
Firstly,  through the 2-body density
$\rho_2(0, {\scriptstyle  \frac{M_{23}}{M_{234}}}\vec{\Delta})$ 
of eq.(\ref{rho2rc}). 
In the limit of a infinite massive core, 
${\lim}_{m_4 \rightarrow \infty}
\rho_2(0, {\scriptstyle  \frac{M_{23}}{M_{234}}} \vec{\Delta}) =  1$,
but this is a poor approximation in the cases considered.
Secondly, the halo wave function is involved through the 2--body density 
$\rho_{2}(\frac{m_3}{M_{23}}\vec{q}+\frac{m_3}{M_{234}}\vec{\Delta},\, \vec{q})$ in eq.(\ref{t12t14li11}). In the limit of a infinite massive core,
${\lim}_{m_4 \rightarrow \infty}
\rho_{2}(\frac{m_3}{M_{23}}\vec{q}+\frac{m_3}{M_{234}}\vec{\Delta},\,
 \vec{q}) =  \rho_v(\vec{\Delta})$, 
and this limit was found to be a good approximation here.

There are several consequences which flow from our analysis. 

We have  shown  that core recoil effects are important.
The same claim has been made by others but within the framework of formalisms 
which differ from ours.
\cite{recoil1} corrects the projectile matter density as a whole for recoil 
effects. One of our points is that
we find no justification  for describing the scattering of protons 
from a light system such as $^{11}$Li in terms of an optical potential 
expressed as a
nucleon-nucleon transition amplitude, $t_{NN}$, 
and a total matter density given by the sum of the valence density
and a core density modulated by a centre-of-mass factor
$ \rho_2(0, {\scriptstyle  \frac{M_{23}}{M_{234}}} \vec{\Delta})$. 
In the first 
place even in the 1st order term it is the N-core transition
amplitude which is modulated in this way. Secondly centre-of-mass 
corrections to the second order terms do not have the structure that would 
arise from iterating the 1st order term as would be expected in a $"t\rho"$
type optical model theory. Eqs. (\ref{t12}),(\ref{t14li11}) and
(\ref{t12t14li11}) can be made to have this structure if the following 3
assumptions are made:

    (i) $t_{14}$ is approximated by its  $"t\rho"$ limit.

    (ii) The average $t_{NN}$ matrices for the core and halo nucleons
         are assumed equal.

    (iii) The limit $m_4 \rightarrow \infty$ for the core mass is assumed.

In our calculations we can find no justification for (i) and the
inadequacy of (ii) was shown very clearly in \cite{CJT1,Garcilazo}.
We have shown here that (iii) is a poor approximation in eq.(\ref{rho2rc}).
In \cite{recoil2} core recoil effects are taken into account in a way which is 
consistent with a few-body model of the reaction and without making a multiple 
scattering expansion. It is, however, not as transparent as in our formalism
how the halo density functions contribute to the scattering. 
An advantage of our approach is that  reaction mechanism and structure
effects are clearly delineated.

\section{Conclusions}
 
 We have seen in this work that 2-body correlation effects associated with
the double scattering term are small in the case of $^{11}$Li scattering
from a proton target at 800MeV/nucleon. The density distribution of the core centre of mass, $\rho_2(0, {\scriptstyle  \frac{M_{23}}{M_{234}}} \vec{\Delta})$, does, however, have a large effect on the calculated crosssection.

 We have shown that the halo structure information associated with  $\rho_{v}(\Delta)$ and 
 $\rho_2(0, {\scriptstyle  \frac{M_{23}}{M_{234}}} \vec{\Delta})$
does not contribute to the scattering simply combined as a total matter density.
Thus, a  proper treatment of the reaction mechanism for 
halo nuclei elastic scattering needs necessarily to incorporate 
structure   features that go beyond  knowledge of 
the total halo matter density distribution alone. 

In summary, we advocate that in microscopic theories of proton scattering 
from light nuclei such as halo nuclei, 
at intermediate and high energies the multiple scattering 
series for the full transition amplitude should be used and that the optical
 potential concept should be avoided.

\bigskip \bigskip

{\bf Acknowledgements:}

This work is supported by Funda\c c\~ao para a Ci\^encia e Tecnologia
(Portugal) through grant No. PRAXISXXI/PCEX/P/FIS/4/96, and by EPSRC(UK) 
through Grant No. GR/J95867.
We would like to thank Professor Ian Thompson for  providing us with the $^{11}$Li
wave functions in a convenient form.

\pagebreak

\newpage

\begin{figure}[h!]
\centerline{
	\parbox[t]{2.5in}{
	\epsfig{file=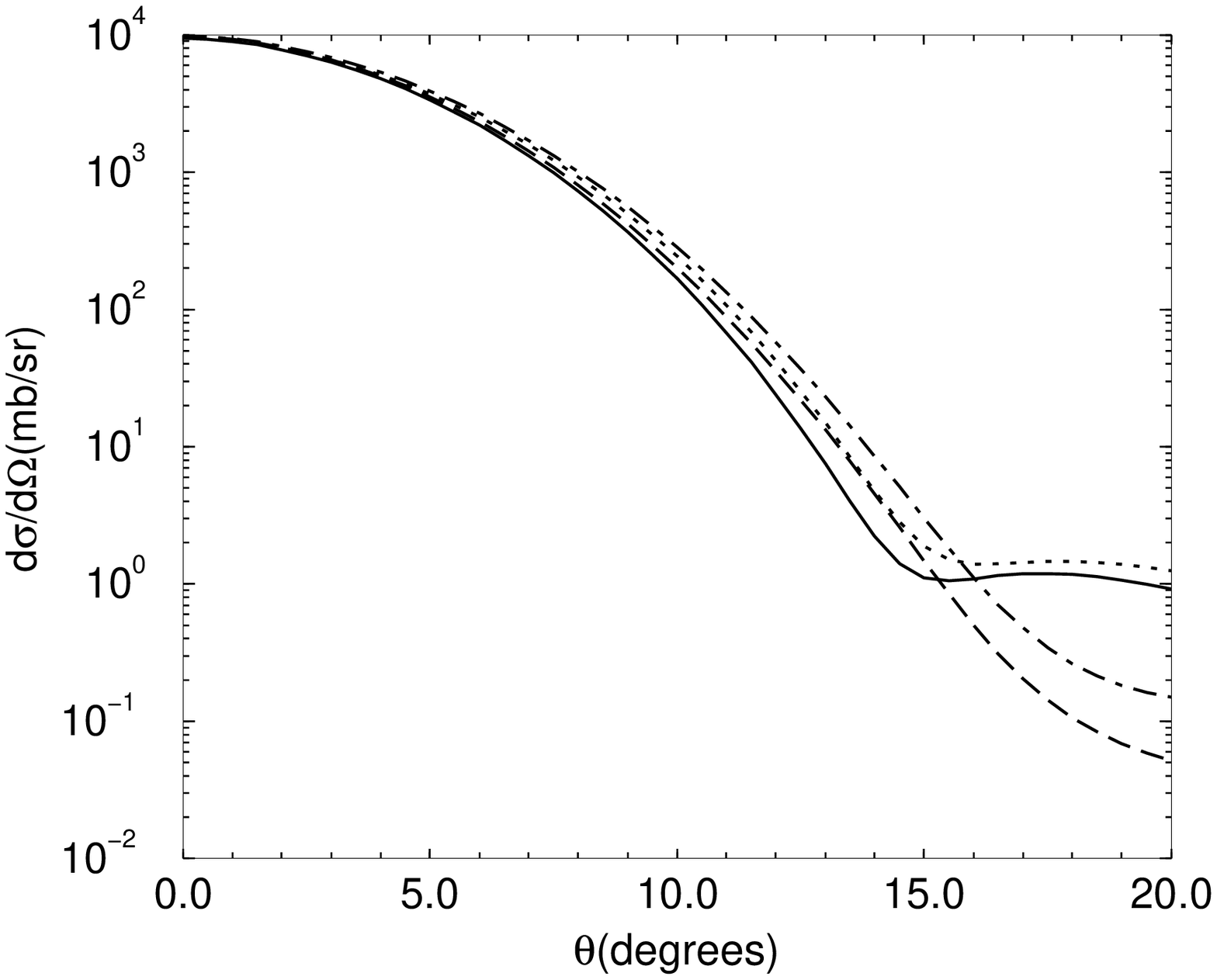,height=3.5in,width=3.5in}}
}
	\caption{ Differential cross section for proton scattering from
$^{11}$Li at 800MeV/nucleon.
The dashed curve was evaluated from the single scattering contributions.
The solid curve includes in addition the double scattering contributions. 
The other curves in the figure are obtained by ignoring   
the relative motion of the core and projectile centres of mass
in the single scattering term and the 
dotted-dashed and dotted curves are cross sections calculated without and 
with double scattering terms respectively.}
	\label{fig:fig}
\end{figure}

\end{document}